\documentclass[12pt]{article}
\usepackage[dvips]{graphicx}
\usepackage{latexsym}
\usepackage{amssymb}

\newcommand{\C}{\mathbb{C}}
\newcommand{\CP}{\mathbb{CP}}

\newcommand{\R}{\mathbb{R}}

\def\ov{\overline}

\def\p{\partial}

\newcommand{\bea}{\begin{eqnarray}}
\newcommand{\eea}{\end{eqnarray}}

\renewcommand{\d}{\mathrm{d}}

\newcommand{\koniec}{\begin{flushright}  $\Box $ \end{flushright}}
\def\be{\begin{equation}}
\def\ee{\end{equation}}

\def\theequation{\thesection.\arabic{equation}}

\def\Om{\Omega}

\def\om{\omega}

\def\ov{\overline}

\def\p{\partial}

\def\ov{\overline}

\newcommand{\hook}{{\setlength{\unitlength}{11pt}   % adjust pt size here
                   \begin{picture}(.833,.8)
                   \put(.15,.08){\line(1,0){.35}}
                   \put(.5,.08){\line(0,1){.5}}
                   \end{picture}}}

\newtheorem{theo}{Theorem}[section] 
\newtheorem{prop}[theo]{Proposition}  
\newtheorem{lemma}[theo]{Lemma}

\newcommand{\A}{\mathbb{A}}
\newcommand{\F}{\mathbb{F}}

\newcommand{\ID}{\; \mbox{$1 \hspace{-1.0mm}  {\bf l}$}_{2} \;}

\newcounter{mnotecount}[section]

\renewcommand{\themnotecount}{\thesection.\arabic{mnotecount}}

\newcommand{\mnote}[1]%{}%
{\protect{\stepcounter{mnotecount}}$^{\mbox{\footnotesize
$%\!\!\!\!\!\!\,
\bullet$\themnotecount}}$ \marginpar{%\color{red}%
\raggedright\tiny\em
$\!\!\!\!\!\!\,\bullet$\themnotecount: #1} }

\begin{document}
\title{\vskip -70pt
\begin{flushright}
{\normalsize DAMTP-2011-75} \\
\end{flushright}
\vskip 10pt
{\bf $SU(2)$ solutions to self--duality equations in eight dimensions 
\vskip 15pt}}
\author{Maciej Dunajski\thanks{Email: M.Dunajski@damtp.cam.ac.uk}
\,\,and\,\, Moritz Hoegner\thanks{Email: M.Hoegner@damtp.cam.ac.uk}
\\
Department of Applied Mathematics and Theoretical Physics,\\
University of Cambridge,\\
Wilberforce Road, Cambridge CB3 0WA, UK. }
\date{}
\maketitle
\begin{center}
{\em Dedicated to Jerzy Lukierski on the occasion of his 75th birthday.}
\end{center}
\begin{abstract}
We consider the octonionic self--duality equations on  eight--dimensional
 manifolds of the form $M_8=M_4\times \R^4$, where
$M_4$ is a hyper--K\"ahler four--manifold. 
We construct explicit solutions to these equations and their symmetry reductions to the non--abelian Seiberg--Witten equations on $M_4$ in the case
when the gauge group is $SU(2)$. These solutions are singular
for flat and Eguchi--Hanson backgrounds. 
For $M_4=\R\times {\mathcal G}$ with a cohomogeneity
one hyper--K\"ahler metric, where ${\mathcal G}$ is a nilpotent (Bianchi II) Lie group, we find a solution which is singular only on a single--sided domain wall. This gives rise to a regular solution of the non--abelian Seiberg--Witten equations on a four--dimensional nilpotent Lie group which carries a regular conformally hyper--K\"ahler metric.
\end{abstract}
%%%%%%%%%%%%%%%%%%%%%%%%%%%%%%%%%%%%
\section{Introduction} 
Gauge theory in dimension higher than four  
has been  investigated in both theoretical physics 
\cite{Corrigan:1982th,Fubini:1985jm,strominger,Baulieu:1997jx,hall} 
and pure  mathematics  \cite{DonT96,Tian} contexts. While the solutions to the 
full second
order Yang--Mills equations seem to be out of reach, the first order 
higher  dimensional analogues of four--dimensional  self--duality equations admit some explicit solutions.
Such equations can be written down on any $n$--dimensional Riemannian 
manifold $M_n$,
once a closed differential form $\Omega$ of degree $(n-4)$ has been chosen.
The generalised self--duality equations state
that the curvature two--form of a Yang--Mills connection takes
its values in one of the eigenspaces of the linear
operator $T:\Lambda^2(M_n)\rightarrow \Lambda^2(M_n)$ given by
$T(\F)=*(\Omega\wedge \F)$. The full Yang--Mills equations are then implied by the Bianchi identity. If $n=4$, and the zero--form 
$\Omega=1$ is canonically given by
the orientation, the eigen-spaces of $T$ are both two--dimensional, and 
are interchanged by reversing the orientation. In general the 
eigen-spaces corresponding to different eigenvalues have different  
dimensions.
For the construction to work, one of these eigen-spaces must
have dimension equal to $(n-1)(n-2)/2$, as only then
the number of equations matches the number of unknowns modulo gauge.

Any Riemannian manifold with special holonomy $Hol\subset SO(n)$ 
admits a preferred parallel $(n-4)$--form,
and the eigen--space conditions above can be equivalently stated
as $\F\in \mathfrak{hol}$, where we have identified 
the Lie algebra $\mathfrak{hol}$ of the holonomy group with a subspace
of $\Lambda^2(M_n)\cong\mathfrak{so}(n)$.
One of the most interesting cases corresponds to eight--dimensional 
manifolds with holonomy $Spin(7)$. The only currently known explicit 
solution 
on $M_8=\R^8$ with its flat metric has a gauge group $Spin(7)$. 
The aim of this paper is to construct explicit solutions 
to the system 
\[
*_8(\F\wedge\Om)=-\F,
\] 
with gauge group $SU(2)$. This  will be achieved by  exploiting the embedding
$SU(2)\times SU(2)\subset Spin(7)$. This holonomy reduction
allows a canonical symmetry reduction to the Yang--Mills--Higgs system in 
four dimensions-- a non--abelian analogue of the
Seiberg--Witten equations involving four Higgs fields 
\cite{DonT96, Baulieu:1997jx, hadys}. 
The explicit $SU(2)$
solutions arise from a t'Hooft-like ansatz which turns out to be 
consistent despite a vast overdeterminancy of the equations.
The resulting solutions  on $\R^8$ fall into two classes, both of which are
singular along a hypersurface. To overcome this, and to evade  Derrick's
theorem prohibiting finite action solutions in dimensions higher than four
we shall consider the case of curved backgrounds of the form
$M_8=M_4\times \R^4$, where $M_4$ is hyper--K\"ahler. The gauge
fields on  the Eguchi--Hanson gravitational instanton are
still singular, but if $M_4$ is taken to be a Bianchi II gravitational 
instanton representing a domain wall \cite{Gibbons:1998ie}, then the Yang--Mills curvature
is regular away from the wall. This gives rise to a regular solution of the non--abelian Seiberg--Witten equation on a four--dimensional nilpotent Lie group ${\mathcal H}$ which carries a regular conformally hyper--K\"ahler metric.
\begin{theo}
\label{theo1}
Let  ${\mathcal H}$ be the simply--connected Lie group whose
left--invariant one--forms satisfy the Maurer--Cartan relations
\[
d\sigma_0=2\sigma_0\wedge\sigma_3-\sigma_1\wedge\sigma_2, \quad d\sigma_1=\sigma_1\wedge\sigma_3, 
\quad d\sigma_2=\sigma_2\wedge\sigma_3,\quad
d\sigma_3=0.
\]
\begin{itemize}
\item The left--invariant metric  
$
\hat{g}=
{\sigma_0}^2+{\sigma_1}^2+{\sigma_2}^2+{\sigma_3}^2
$
on ${\mathcal H}$
is regular and conformally hyper--K\"ahler.
\item The $\mathfrak{su}(2)$--valued one--forms
\[
A=\frac{3}{4}(\sigma_2\otimes T_1-\sigma_1\otimes T_2+\sigma_0\otimes T_3),\quad \Phi=-\frac{\sqrt{21}}{3} A
\]
with $
[T_1, T_2]=T_3, \; [T_3, T_1]=T_2, \; [T_2, T_3]=T_1
$
satisfy
\[
F_+ = \frac{1}{2} [\Phi, \Phi]_+,\quad 
(D\Phi)_- =0, \quad D \ast_4 \Phi =0,
\]
where $D=d+[A, \dots], F=dA+A\wedge A$, and $\pm$ denote self--dual
(+) and anti--self--dual (-) parts with respect to $\hat{g}$.
\end{itemize}
\end{theo} 
 Finally we should mention that there
are other candidates for `self--duality' equations in higher dimensions.
One possibility in dimension eight, exploited by Polchinski in the context
of heterotic string theory \cite{Pol}, is to consider the system
$
*\F\wedge \F=\pm \F\wedge \F.
$
These equations are conformally invariant, and thus the finite action solutions
compactify $\R^8$ to the eight-dimensional sphere, but unlike the system
(\ref{asdym}) considered in this paper
they do not imply the Yang--Mills equations.
\subsubsection*{Acknowledgements}
We thank Gary Gibbons, Hermann Nicolai and Martin Wolf 
for for useful discussions.
%%%%%%%%%%%%%%%%%%%%%%%%%%%%%%%%%%%%%%%%%%%%%%%%%%%%%%%%%%%%%%%%%%%%%%%%%%%%%%%%%%%%%%%%%%%%%
\section{Self--duality in eight dimensions}

Let $(M_8, g_8)$ be an eight--dimensional oriented Riemannian manifold.
The 21--dimensional Lie group $Spin(7)$ is subgroup of $SO(8)$ preserving a self--dual
four--form $\Omega$.
Set  $e^{\mu \nu \rho \sigma} = e^{\mu} \wedge e^{\nu} \wedge e^{\rho} \wedge e^{\sigma}$. There exists an orthonormal frame in which 
the metric and the four--form
are given by
\begin{eqnarray}
\label{Omeganull}
g_8&=&(e^0)^2+(e^1)^2+\dots+(e^7)^2,\nonumber\\
\Omega &=& e^{0123} + e^{0145} + e^{0167} + e^{0246} - e^{0257} - e^{0347} - e^{0356}\\
 \nonumber         &&- e^{1247} - e^{1256} - e^{1346} + e^{1357} + e^{2345} + e^{2367} + e^{4567}.
\end{eqnarray}
Let $T:\Lambda^2(M_8)\rightarrow \Lambda^2(M_8)$ be a self-adjoint operator
given by 
\[\om\rightarrow *_8(\Om\wedge\om),\]
where $*_8$ is the Hodge operator of $g_8$ corresponding to the orientation
$\Omega\wedge\Omega$.
The $28$--dimensional space of two-forms in eight dimensions splits
into $\Lambda^2_{21}\oplus \Lambda^2_+$, where $\Lambda^2_{21}$ and
$\Lambda^2_+$ are eigenspaces of $T$ with eigenvalues $-1$ and $3$ respectively.
The $21$--dimensional space $\Lambda^2_{21}$ can be identified with the 
Lie algebra $\mathfrak{spin}(7)\subset\mathfrak{so}(8)\cong \Lambda^2(M_8)$.

Let $\A$ be a one--form on $\R^8$ with values in a Lie algebra
$\mathfrak{g}$ of a gauge group $G$.
The $Spin(7)$  self--duality  condition states that the curvature two form 
\[
\F=d\A+\frac{1}{2}[\A, \A]
\]
takes its values in $\Lambda^2_{21}$. This leads to a system of seven first order equations\
\be
\label{asdym}
*_8(\F\wedge\Om)=-\F,
\ee
explicitly given by\newpage
\begin{eqnarray*}
\nonumber
\F_{01} + \F_{23} + \F_{45} + \F_{67} &=& 0,\\
 \nonumber
 \F_{02} - \F_{13} + \F_{46} - \F_{57} &=& 0,\\
 \nonumber
 \F_{03} + \F_{12} - \F_{47} -\F_{56} &=& 0,\\
 \F_{05} + \F_{14} + \F_{27} + \F_{36} &= &0,\\
 \nonumber
 \F_{06} - \F_{17} + \F_{24} - \F_{35} &=& 0,\\
 \nonumber
 \F_{07} + \F_{16} - \F_{25} - \F_{34} &=& 0,\\
  \nonumber
 \F_{04} - \F_{15} - \F_{26} + \F_{37} &=& 0.
\end{eqnarray*}
This is a determined system of PDEs as one of the eight components of $\A$
can be set to zero by a gauge transformation
\[
\A\longrightarrow  \rho\A \rho^{-1} -d\rho\; \rho^{-1}, \quad\mbox{where}\quad
\rho\in \mbox{Map}(M_8, G).
\]
Equations (\ref{asdym}) were first investigated in \cite{Corrigan:1982th},  and some solutions  
were found in \cite{Fairlie:1984mp,Fubini:1985jm} for the gauge group 
$Spin(7)$. 
If $\A$ is a solution to (\ref{asdym}), then it is 
a Yang-Mills connection because
\[
D*_8\F=-D\F\wedge\Om=0, \quad \mbox{where}\quad D=d+[\A, \dots]
\]
by the Bianchi identities.\footnote{The Derrick 
scaling argument (see e.g. \cite{Dbook}) shows
there are no nontrivial finite
action solutions to  the pure Yang--Mills equations on $\R^8$. 
This obstruction can be overcome if some dimensions are
compactified. 
If $(M_8, g_8)$ is a compact manifold with holonomy $Spin(7)$, then
the YM connections which satisfy (\ref{asdym})
are absolute minima of the Yang--Mills functional
\[
E(\A)=\frac{1}{4\pi}\int_{M_8}|\F|^2\; \mbox{vol}_{M_8}.
\]
To see this write $\F=\F_++\F_-$, where $\F_+\in \Lambda^2_+, 
\F_-\in\Lambda^2_{21}$, and verify that
\[
\F\wedge *_8\F=\F_+\wedge *_8\F_+ +\Om\wedge \F\wedge \F.
\]
The integral of the trace 
of the second term on the RHS  is independent on $\A$.}
%%%%%%%%%%%%%%%%%%%%%%%%%%%%%%%%%%%%%%%%%%%%%%%%%%%%%%%%%%%%%%%%%%
\subsection{Non--abelian Seiberg--Witten equations}
\subsubsection*{Holonomy reduction}
Equations (\ref{asdym}) are valid on curved eight--dimensional 
Riemannian manifolds with holonomy equal to, or contained in $Spin(7)$,
as such manifolds are characterised by the existence of
a parallel four--form given by (\ref{Omeganull}). We shall consider
the special case of product manifolds \cite{Joyce}
\be
\label{product}
M_8=M_4\times \widetilde{M}_4, \quad g_8=g_4+\tilde{g}_4,
\ee
where $M_4$ and $\widetilde{M}_4$  are hyper--K\"ahler manifolds.
Let ${\psi_i}^{\pm}$ span the spaces  $\Lambda^2_+(M_4)$ and
$\Lambda^2_-(M_4)$ of
self--dual and anti--self--dual  
two--forms respectively.
 Thus
\be
\label{two_f}
g_4=(e^0)^2+(e^1)^2 +(e^2)^2 +(e^3)^2, 
\quad\mbox{and}\quad{\psi_i}^{\pm}=e^0\wedge e^i\pm
\frac{1}{2}{\varepsilon^{i}}_{jk}e^j\wedge e^k, 
\ee
where $i, j, \dots =1, 2, 3$ with
analogous expressions for $\tilde{g}_4$.
The $Spin(7)$ four--form (\ref{Omeganull})
is then given by
\[
\Omega=\mbox{vol}+\widetilde{\mbox{vol}}+\sum_{i,j=1}^3
\eta_{ij}{\psi_i}^+\wedge{\tilde{\psi}_j}^+,
\]
where $\eta=\mbox{diag}(1, 1, -1)$ and vol, $\widetilde{\mbox{vol}}$ are volume forms on $M_4$ and $\widetilde{M}_4$ respectively. 
The self--dual four--form $\Omega$ is closed as a consequence
of the closure of $\psi_i$ and $\tilde{\psi}_i$ which can always be achieved
by a choice of the orthonormal frame on hyper--K\"ahler manifolds.

\subsubsection*{Symmetry reduction}
We shall now consider the self--duality equations (\ref{asdym}) for 
a $\mathfrak{g}$-valued connection $\A$ over an eight-manifold 
$M_8$ of the form (\ref{product}), where $M_{4}$ is an arbitrary 
Hyper-K\"ahler four manifold, and $\widetilde{M}_4=\widetilde{\R}^4$ 
is flat.
We shall look for solutions $\A$ that 
admit a four-dimensional symmetry group generated by the translations 
on $\widetilde{\R}^4$. If $x^{\mu}$ are local coordinates of $M_8$, then we denote the coordinates of $M_4$ by $x^a$ and those of $\tilde{\R}^4$ by $\tilde{x}^a$.  The Greek indices run from 0 to 7 as Latin indices run from 0 to 3.
We choose a frame $e^{\mu}$ in (\ref{Omeganull}),/
where $e^{\mu} \; (\mu = 0,\;\dots\;,3)$ is a  frame (\ref{two_f})
on $M_{4}$ in which $\psi_i$ are closed and 
$e^{\mu}=d\tilde{x}^{\mu-4} \; (\mu = 4,\;\dots\;,7)$. 
We can then write
\begin{eqnarray}
\label{Asym}
\nonumber
 \mathbb{A} 
&=& \sum_{\mu=0}^7\A_{\mu} (x^b) e^{\mu} \\
&=& \sum_{a=0}^3A_a(x^b) e^a + \Phi_0(x^b) e^4 - \Phi_1(x^b) e^5 - \Phi_2(x^b) e^6 + \Phi_3(x^b) e^7\\
\nonumber
&=& A + \Phi'
\end{eqnarray}
where we have re--labelled coefficients and consequently defined $A$, $A_a$, $\Phi'$ and $\Phi_a$. Thus $A$ is a $\mathfrak{g}$--connection on $M_4$. Let $F$ denote the 
curvature of $A$, and let $F_\pm$ be the SD and ASD parts of $F$
with respect to the Hodge operator $\ast_4$ of $g_4$.
Furthermore, we introduce the following notation:
Let $\Phi=\Phi_a e^a$ be a $\mathfrak{g}$--valued one--form and let
$\nabla_a$ be four vector fields dual to $e^a$, i. e. 
$\nabla_a\hook e^b=\delta^b_a$. Set
$D = e^a\otimes\nabla_a + \left[ A, \cdot \right]$, and
$D_a \Phi_b = \partial_a \Phi_b + [A_a, \Phi_b]$. Thus $D \Phi = D_{\left[ a \right.}\Phi_{\left. b \right]} e^a \wedge e^b$ captures the antisymmetric part of $D_a \Phi_b$. Note that $A$, $F$, $\Phi$ and $D \Phi$ are $\mathfrak{su}(2)$-valued forms over $M_{4}$. We are thus splitting up the connection and curvature in various pieces. Note that $\Phi' \neq \Phi_a e^a$ due to the choice of indices and signs in (\ref{Asym}).

 Now we shall investigate the  equations (\ref{asdym}) on the chosen product background $M_8$. Invoking translational symmetry along $\tilde{\R}^4$ as explained, we find the following
\begin{prop}
For a connection of the form {\em(\ref{Asym})} equations {\em(\ref{asdym})} 
reduce to the following system of equations for the differential forms $A$ and $\Phi$ over $M_{4}$:
\begin{eqnarray}
\label{F+=0}
\label{FPhi} F_+ - \frac{1}{2} [\Phi, \Phi]_+ &=&0 \\
 \label{DPhi} [D\Phi]_- &=&0 \\
 \label{DivPhi} D \ast_4 \Phi &=&0,
\end{eqnarray}
where the $\pm$  denote  the SD (+) or ASD (-) part with respect 
to the Hodge operator $\ast_4$. 
\end{prop}
{\bf Proof.}
This reduction  has been performed before \cite{Baulieu:1997jx,DonT96,klemm,hadys}, but in the slightly 
different context\footnote{In the approach of \cite{hadys} $M_8$ is the total space of the spinor bundle over $M_4$ and equations 
(\ref{DPhi}) and (\ref{DivPhi}) are combined into the 
non--abelian Dirac equation}. 
We shall present a proof adapted to our setup.
One obtains these equations by inserting the explicit expression for $\A = A + \Phi'$ and the definition of the curvature, $\F=d\A + \frac{1}{2} [\A , \A]$ into the system (\ref{asdym}). For the curvature, we find
\begin{eqnarray*}
 \F &=& d\A + \frac{1}{2} [\A , \A] \\
    &=& dA + d\Phi' + \frac{1}{2} [ A , A ] + [ A , \Phi' ] + \frac{1}{2} [ \Phi', \Phi' ] \\
    &=& F + D\Phi' + \frac{1}{2} [ \Phi' , \Phi' ].
\end{eqnarray*}
In the expression $\F = \frac{1}{2} \F_{\mu \nu} e^{\mu} \wedge e^{\nu}$, the two--form 
$F$ accounts for coefficients $\F_{\mu \nu}$ with both indices in the range $0 \leq \mu, \nu \leq 3$, the term $\frac{1}{2} [ \Phi', \Phi']$ for those coefficients $\F_{\mu \nu}$ with indices in the range $4 \leq \mu, \nu \leq 7$ and $D\Phi'$ for coefficients with one index each. This allows us to translate the components $\F_{\mu \nu}$, e.g.
\[
 \F_{01} = F_{01}, \quad\F_{25}= \left( D\Phi' \right)_{25}  = -  D_2 \Phi_1, 
 \quad \F_{67}  = \frac{1}{2} \left[ \Phi' , \Phi' \right]_{67} = - \frac{1}{2} \left[ \Phi_2 , \Phi_3 \right].
\]
The sign and index changes are a result of the labelling of the 
components of $\Phi'$. Applying this to the system (\ref{asdym}), we find
\begin{eqnarray*}
 F_{01} + F_{23} - \frac{1}{2} \left[ \Phi_0 , \Phi_1 \right] - \frac{1}{2} \left[ \Phi_2 , \Phi_3 \right] &=& 0,\\
 F_{02} - F_{13} - \frac{1}{2} \left[ \Phi_0 , \Phi_2 \right] + \frac{1}{2} \left[ \Phi_1 , \Phi_3 \right] &=& 0,\\
 F_{03} + F_{12} - \frac{1}{2} \left[ \Phi_0 , \Phi_3 \right] - \frac{1}{2} \left[ \Phi_1 , \Phi_2 \right] &=& 0,\\
 - D_0 \Phi_1 + D_1 \Phi_0 + D_2 \Phi_3 - D_3 \Phi_2 &=& 0,\\
 - D_0 \Phi_2 - D_1 \Phi_3 + D_2 \Phi_0 + D_3 \Phi_1 &=& 0,\\
   D_0 \Phi_3 - D_1 \Phi_2 + D_2 \Phi_1 - D_3 \Phi_0 &=& 0,\\
 D_0 \Phi_0 + D_1 \Phi_1 + D_2 \Phi_2 + D_3 \Phi_3 &=& 0.
\end{eqnarray*}
This is exactly the system (\ref{F+=0}) with all components written out. \koniec
The resulting system is  set of equations for a connection $A$ and four non-abelian Higgs fields $\Phi_a$ over $M_{4}$. In particular they can be regarded as a non-abelian version \cite{Baulieu:1997jx,DonT96,popov,klemm,hadys} of the equations found by Seiberg and Witten 
\cite{Seiberg:1994rs}. We will call (\ref{F+=0}) the non-abelian Seiberg--Witten equations.
%%%%%%%%%%%%%%%%%%%%%%%%%%%%%%%%%%%%%%%5
\section{Ansatz for $SU(2)$ solutions}
To find explicit solutions to (\ref{F+=0}) and (\ref{asdym}) with 
the gauge group $SU(2)$
we shall proceed with an analogy to the t'Hooft
ansatz for the self--dual Yang--Mills equations on $\R^4$. 

Let $T_i, (i = 1, 2, 3)$ denote a basis of $\mathfrak{su}(2)$ with commutation relations $[T_i, T_j]=\epsilon_{ijk} T_k$ and 
$T_i T^i:=T_iT_j\delta^{ij} = - \frac{3}{4} \ID $. We can then define 
two $\mathfrak{su}(2)$--valued
two--forms $\sigma$ and $\tilde{\sigma}$ such that $*_4\sigma=\sigma$ and $*_4\tilde{\sigma}=-\tilde{\sigma}$
by
\be
\label{thoft}
 \sigma = \frac{1}{2} \sigma_{ab} e^a \wedge e^b = \sum\limits_i T_i \; 
{\psi_i}^+,\quad
 \tilde{\sigma} = \frac{1}{2} \tilde{\sigma}_{ab} e^a \wedge e^b = \sum\limits_i T_i \; {\psi_i}^-, 
\ee
where ${\psi_i}^{\pm}$ are given by (\ref{two_f}). 
Thus the forms $\sigma_{ab}$
select the three--dimensional space of SD two forms $\Lambda^2_+(M_4)$
from the six--dimensional space $\Lambda^2(M_4)$ and project
it onto the three--dimensional 
subspace $\mathfrak{su}(2)$ of $\mathfrak{so}(4)$.
An analogous isomorphism between $\Lambda^2_-(M_4)$ and another copy of 
$\mathfrak{su}(2)$ is provided by $\tilde{\sigma}$.
The following identities hold
\be
\label{sigmaId}
\tilde{\sigma}_{ab} \sigma^{ab} = 0,\quad
\sigma_{ab} {\sigma^b}_c = \frac{3}{4} \; \mbox{$1 \hspace{-1.0mm}  {\bf l}$}_{2} \; \delta_{ac} + \sigma_{ac}, \quad \sigma_{ab} \sigma^{ab} = -3 \; \mbox{$1 \hspace{-1.0mm}  {\bf l}$}_{2}.
\ee
We now return to equations (\ref{F+=0}) and make the following ansatz
for the $\mathfrak{su}(2)$-valued one-forms $A$ and $\Phi$,
\be
 \label{ansatz}
 A = \ast_4 ( \sigma \wedge dG) = \sigma_{ab} \nabla^b G e^a, \quad
\Phi =  \ast_4( \sigma \wedge dH ) = \sigma_{ab} \nabla^b H e^a,
\ee
where $G, H: M_4 \rightarrow \mathbb{R}$ are functions on $M_4$ and $\nabla_a$ are the vector fields dual to $e^a$. 
Let $\Box = \ast d \ast d + d \ast d \ast$ be the Laplacian and $\nabla$ be the gradient on $M_4$, and let 
$d(e^a) = {C^a}_{bc} e^b \wedge e^c$.  The following Proposition will be proved in the Appendix
\begin{prop}
\label{prop1}
 The non-abelian Seiberg-Witten equations {\em(\ref{F+=0})} 
are satisfied by Ansatz {\em(\ref{ansatz})} if and only if $G$ and $H$ satisfy the following system of coupled partial differential equations:
\begin{eqnarray}
\label{pde1}
 \label{Box1} \Box G + | \nabla G |^2 - | \nabla H |^2 &=&0,\\
 \label{Box12}  \left( {\epsilon_{ea}}^{bc} {C^a}_{bc} \sigma^{ed} - \sigma^{ab} {C^d}_{ab} \right) \nabla_d G &= &0,\\
 \label{PT1} \tilde{\sigma}_{ac} {\sigma^{c}}_{b} \left(\nabla^a \nabla^b H - 2 \nabla^a G \nabla^b H \right) &=&0, \\
 \label{ASD1} \sigma_{ab} \left(\nabla^a \nabla^b H - 2 \nabla^a G \nabla^b H \right) &=&0.
\end{eqnarray}
\end{prop}
Note that equation (\ref{ASD1}) is equivalent to the anti--self--duality
of the  antisymmetric part of
\[
 \nabla^a \nabla^b H - 2 \nabla^a H \nabla^b G.
\]
A similar interpretation of equation (\ref{PT1}) is given by the following
\begin{lemma}
\label{LemPT} Let $\Sigma_{ab}$ be an arbitrary tensor. Then
\begin{equation}
\nonumber
 \tilde{\sigma}^{ab} {\sigma^{c}}_{b} \Sigma_{ac} =0 \quad \Leftrightarrow \quad
 \Sigma_{\left( ac \right)} = \frac{1}{4} {\Sigma_b}^b \delta_{ac}.
 \end{equation}
\end{lemma}
{\bf Proof.} Starting from the left hand side we first define a two--form $(\Sigma \sigma) = {\sigma^c}_{\left[ b \right.} \Sigma_{\left. a \right] c}\; e^a \wedge e^b$. Therefore
\[
  \tilde{\sigma}^{ab} {\sigma^c}_{b} \Sigma_{ac} = \tilde{\sigma}^{ab} {\sigma^c}_{\left[ b \right.} \Sigma_{\left. a \right] c} =  \ast [ \tilde{\sigma} \wedge (\Sigma \sigma) ] = 0,
\]
and so $(\Sigma \sigma)$ is self-dual, i. e.
\begin{equation}
 \nonumber
 (\Sigma \sigma)_{01} = (\Sigma \sigma)_{23}, \quad \quad (\Sigma \sigma)_{02} = - (\Sigma \sigma)_{13}, \quad \quad (\Sigma \sigma)_{03} = (\Sigma \sigma)_{12}.
\end{equation}
 Using the definition (\ref{thoft}) of $\sigma_{ab}$ in terms of the generators of $\mathfrak{su}(2)$ this is equivalent to a system of nine linear equations for the components of $\Sigma_{ac}$:  six of them  set off-diagonal terms to zero, three more equate the four diagonal terms of $\Sigma_{ac}$. Solving this system is straightforward: the only solution is $\Sigma_{\left(ac\right)} = \Sigma \delta_{ac}$ for some scalar function $\Sigma$.  \koniec
Thus equations (\ref{PT1}) and (\ref{ASD1}) together imply that $\nabla^a \nabla^b H - 2 \nabla^a H \nabla^b G$ is the sum of a (symmetric) pure-trace term and an (anti-symmetric) ASD term. To continue with the analysis of (\ref{pde1}) we need to distinguish between flat and curved background spaces.
%%%%%%%%%%%%%%%%%%%%%%%%%%%%%%%%%%%%%%%%%%%%%%%%%%%%%%%%
\subsection{Flat background}
\begin{figure}[ht]
\begin{minipage}[b]{0.5\linewidth}
\centering
\includegraphics[width=50mm]{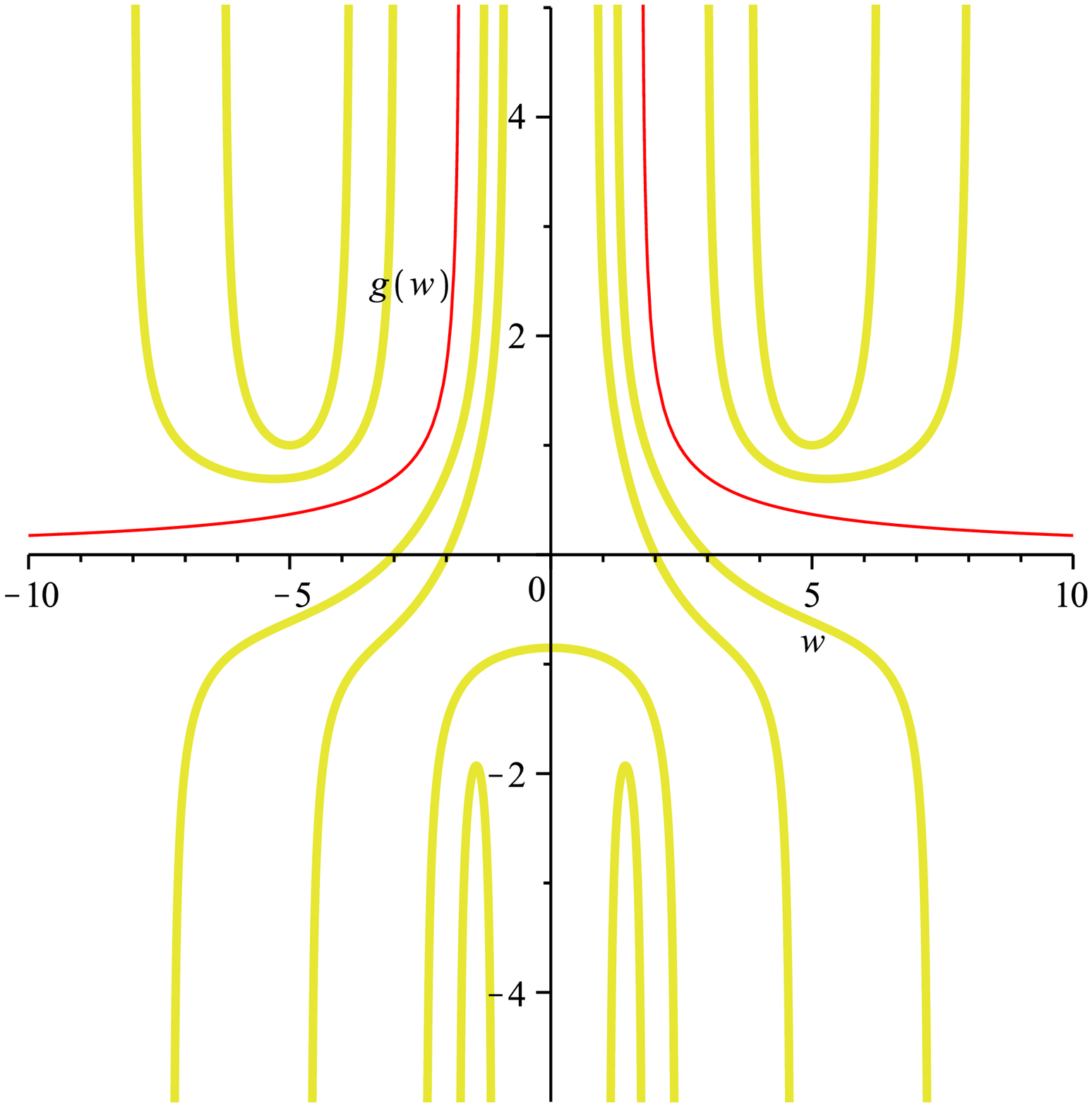}
\caption{Solutions of ODE (\ref{oderr}) I}
\label{R8r1}
\end{minipage}
\hspace{0.5cm}
\begin{minipage}[b]{0.5\linewidth}
\centering
\includegraphics[width=50mm]{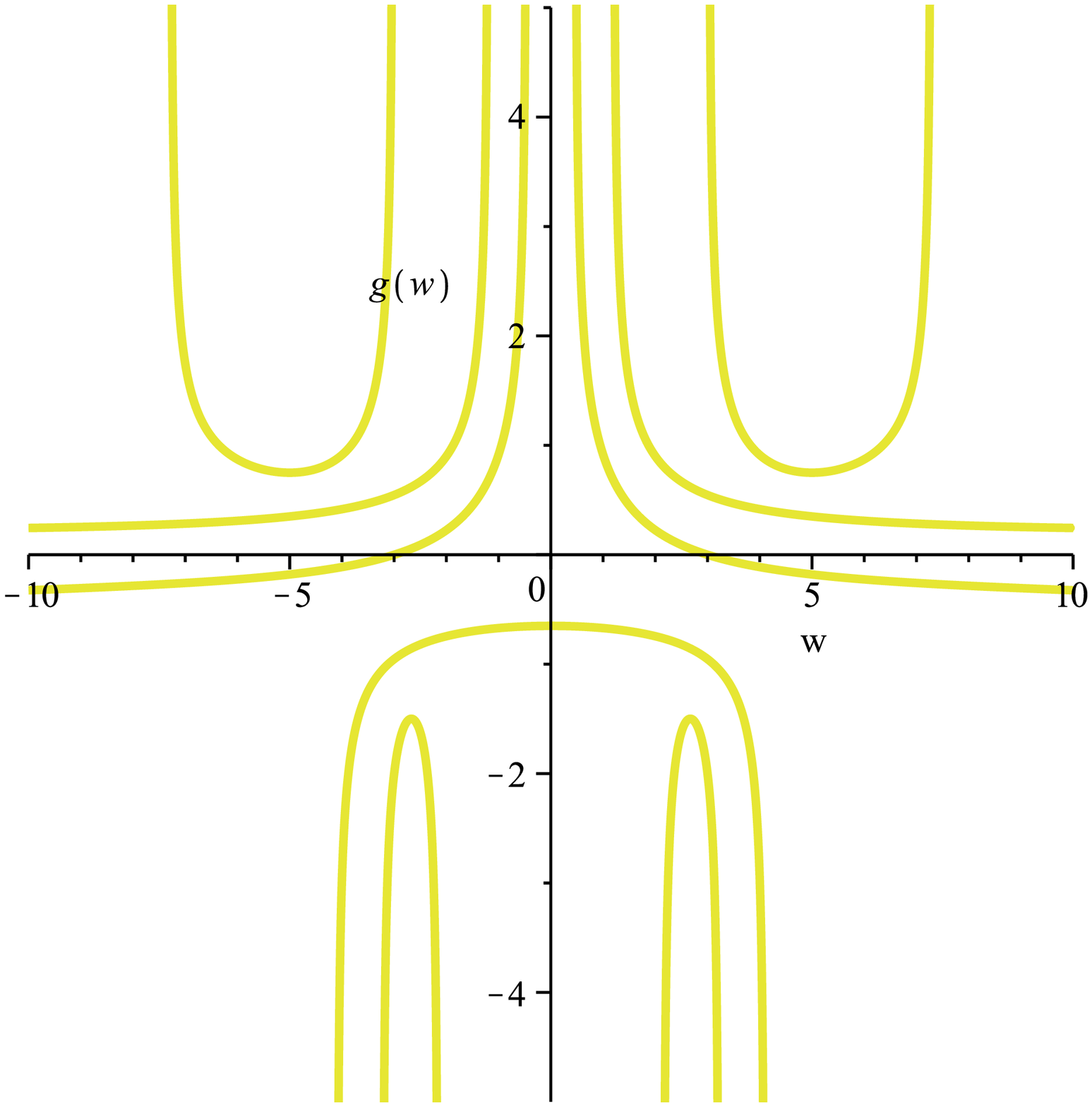}
\caption{Solutions of ODE (\ref{oderr}) II}
\label{R8r2}
\end{minipage}
\end{figure}

Our first choice for $M_4$ is the flat space $\R^4$ with $e^a = dx^a$ for Cartesian coordinates $x^a$. Since the one-forms $e^a$ are closed we have ${C^a}_{bc} = 0$ and the dual vector fields $\nabla_a = \partial_a$ commute. This implies that (\ref{Box12}) is identically satisfied. Equation (\ref{ASD1}) implies that the  simple two-form $dG \wedge dH$ is ASD. Therefore this form is equal to zero, since there are no real simple
 ASD two-forms in Euclidean signature and thus
$H$ and $G$ are functionally dependent. Therefore we can set  $H=H(G)$.
Thus the tensor $\Sigma_{ab}= \partial_a \partial_b H - 2 \partial_a H \partial_b G $ 
is symmetric.
 Next, we turn our attention to (\ref{PT1}). Applying  Lemma \ref{LemPT} we deduce 
that $\Sigma_{ab}$ is pure trace. Defining a one--form $f=\exp{(-2G)}dH$ we 
find that
\be
\label{dafc}
\p_a f_c = \Sigma e^{-2G} \delta_{ac}
\ee
for some $\Sigma$.
Equating the  off-diagonal components of (\ref{dafc}) to zero shows that $f_c$ depends on $x^c$ only, and the remaining four equations
yield
$
dH = e^{2G} dw,
$
where \[w = \frac{1}{2} \gamma x_a x^a + \kappa_a x^a,\] for some constants 
$\gamma, \kappa_a$.
Thus $G$ also depends only on $w$ and, defining $g(w)=\exp{G(w)}$,
equation
(\ref{Box1}) yields
\begin{equation}
\label{ODEF}
g'' (2\gamma w + \kappa^2) + 4\gamma g' - g^5 (2\gamma w + \kappa^2) = 0.
\end{equation}
There are two cases to consider
\begin{itemize}
 \item Assume that $\gamma=0$, in which case
\begin{equation}
\label{oder1st}
 g' = \pm \sqrt{\frac{1}{3} g^6 + \gamma_1}.
\end{equation}
To obtain an explicit solution we set the constant $\gamma_1 = 0$.
Using the translational invariance of (\ref{F+=0}) we can always 
put $w=x^3$. Reabsorbing the constant of integration and rescaling 
yields
\be
\label{static_sol}
G = - \frac{1}{2} \ln|x^3|,  \quad    H = \frac{\sqrt{3}}{2} \ln|x^3|.
\ee
Using these functions in the ansatz (\ref{ansatz}) 
for the pair $(A, \Phi)$ will give rise to a curvature $\F$ such that (\ref{asdym}) holds. Note however that the connection is singular along a hyperplane in $\R^4$ and thus $\A$ is also singular along a hyperplane in $\R^8$ because of the translational symmetry. The curvature for this solution is singular along
a hyper--plane with normal $\kappa_a$, and blows up like $|x^3|^{-2}$, thus the solution is singular.
A numerical plot of solutions of (\ref{oder1st}) for different $\gamma_1$ is displayed in Figure \ref{R8}. Since the equation is autonomous, one can obtain the general solution by translating any curve in the $x^3$-direction. The red line corresponds to (\ref{static_sol}). Note that all other curves have two vertical asymptotes and do not extend to the whole range of $x^3$.
\begin{figure}
\centering
\includegraphics[width=60mm]{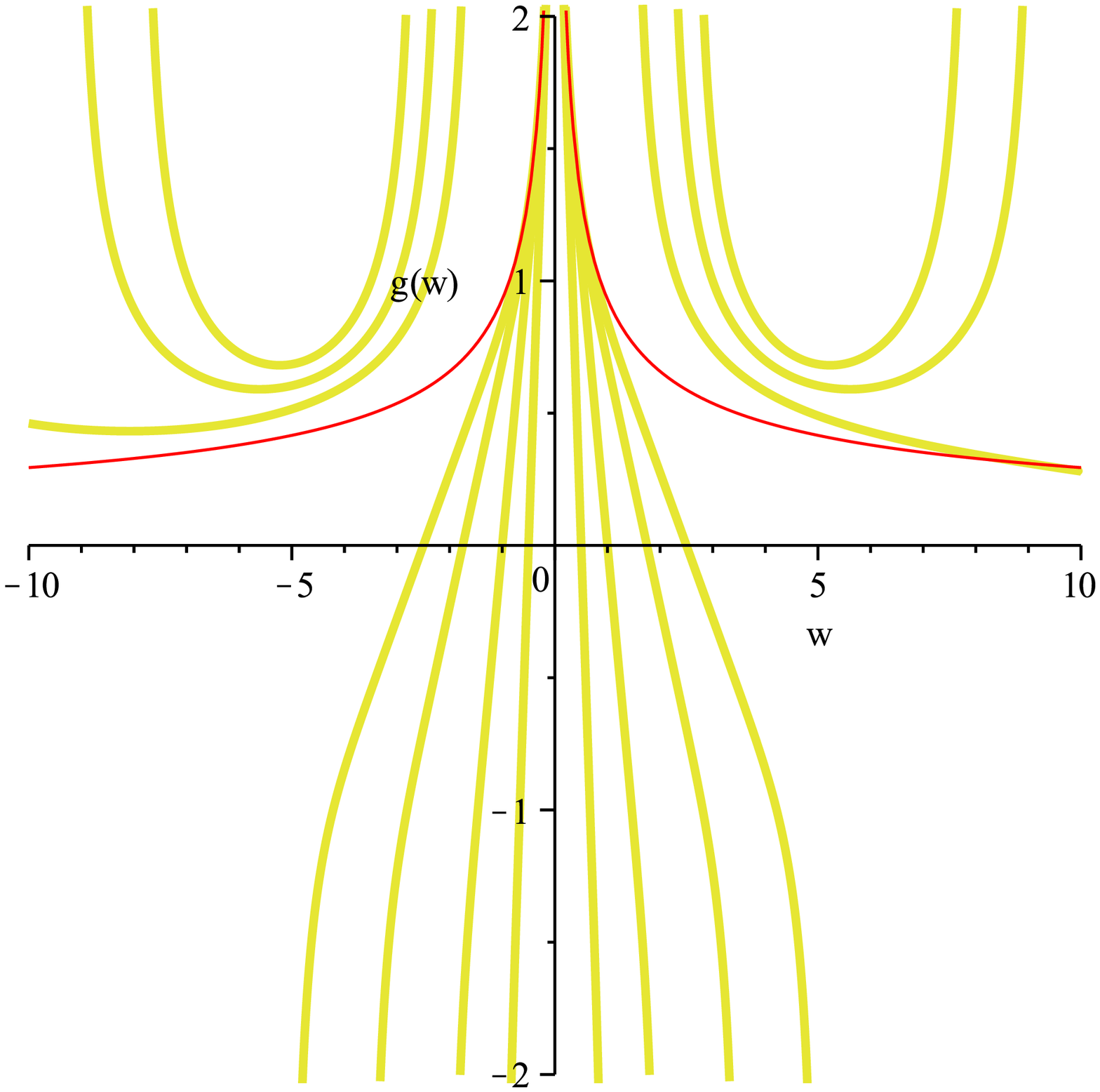}
\caption{Numerical plot of solutions to $g'' = g^5$}
\label{R8}
\end{figure}

%\begin{align*}
% w = \frac{C}{2} \left( x_a + \frac{K_a}{C} \right) \left( x^a + \frac{K^a}{C} \right) - \frac{K^2}{2C}.
%\end{align*}

\item We will now present a second, radially symmetric solution. 
If $\gamma \neq 0$ we translate the independent variable by 
$w\rightarrow  w - \frac{\kappa^2}{2\gamma}$, then (\ref{ODEF}) is

\begin{equation}
 \label{oderr}
 g'' {w} + 2 g' - g^5 {w} =0.
\end{equation}
Figures \ref{R8r1} and \ref{R8r2} contain the numerical plots of two one-parameter families of solutions. An explicit analytic solution is given by
\[
\nonumber
g({w})=\frac{1}{\sqrt{\frac{1}{3} {w}^2 - 1}}.
\]
If we define the radial coordinate $r := \Big| \sqrt{\frac{\gamma}{2 \sqrt{3}}}
 \left( x_a + \frac{\kappa_a}{\gamma} \right) \Big|$, then ${w} = \sqrt{3} r^2$ and
\be
\label{spher_sym}
 G(r)= - \frac{1}{2} \ln \left( r^4 - 1 \right), \quad 
 H(r)= \frac{\sqrt{3}}{2} \ln \left[ \frac{ r^2 - 1}{ r^2 + 1} \right].
\ee
The pair $(A, \Phi)$ in 
(\ref{ansatz}) is singular on the  sphere $r=1$ in $\R^4$. 
In $\R^8$ this corresponds to cylinders of a hypersurface type. 
The curvature is given by
\[
\F = \frac{{K^i}_{\mu\nu} T_i}{(r^4 -1)^2} \; e^\mu \wedge e^\nu,
\]
where ${K^i}_{\mu\nu}$ are quadratic polynomials in $r^2$. The numerical results 
suggest that there are no regular solutions to (\ref{oderr}) and most solution curves do not even extend to the full range of $r$.
\end{itemize}
This concludes the process of solving the initial system of coupled partial differential equations (\ref{pde1}). We have shown that the most general solution to this system is given by two functions of one variable, $G$ and $H$ with $w:=\frac{1}{2} \gamma 
x_a x^a + \kappa_a x^a$, which are determined by an ordinary differential equation. We presented two classes of solutions in closed form.
\subsection{Curved backgrounds}

The solutions we have found in the last subsection 
have extended singularities resulting in an unbounded 
curvature and infinite action. While we could argue
that the former is an artifact resulting from the form
our ansatz, there is no hope to cure the latter. 
The existence of the finite action solutions
to pure Yang--Mills theory on $\R^8$ or to Yang--Mills--Higgs
theory on $\R^4$ is ruled out by the Derrick scaling argument 
\cite{Dbook}.

To evade Derrick's argument
we shall now look at curved hyper--K\"ahler manifolds $M_{4}$ in place of $\R^4$. 
The one-forms $e^a$ in the orthonormal frame (\ref{two_f})
are no longer closed and  the vector fields $\nabla_a$ do not commute, as $C_{ab}^c\neq 0$. The equations (\ref{PT1}) and (\ref{ASD1})  
imply that $\nabla_a \nabla_b H - 2 \nabla_a G \nabla_b H$ is a sum of a pure-trace term and an ASD term, but examining the integrability conditions shows that the trace term vanishes unless the metric $g_4$ is flat. Thus
\begin{equation}
\label{gensol2}
\nabla_a H = \delta_a e^{2G},
\end{equation}
where $\delta_a$ are some constants of integration. We shall analyse
two specific examples of $M_{4}$. The first class of solutions on the Eguchi--Hanson
manifold generalises the spherically symmetric solutions (\ref{spher_sym}),
which were singular at $r=1$. In the Eguchi--Hanson case the parameter
in the metric can be chosen so that $r=1$ does not belong to the manifold.
The second class of solutions on the domain wall backgrounds generalises
the solutions (\ref{static_sol}).
%%%%%%%%%%%%%%%%%%%%%%%%%%%%%%%%%%%%%%%%%%%%%%%%%%%%%%%
\subsubsection*{Eguchi-Hanson background}
\begin{figure}[ht]
\begin{minipage}[b]{0.5\linewidth}
\centering
\includegraphics[width=60mm]{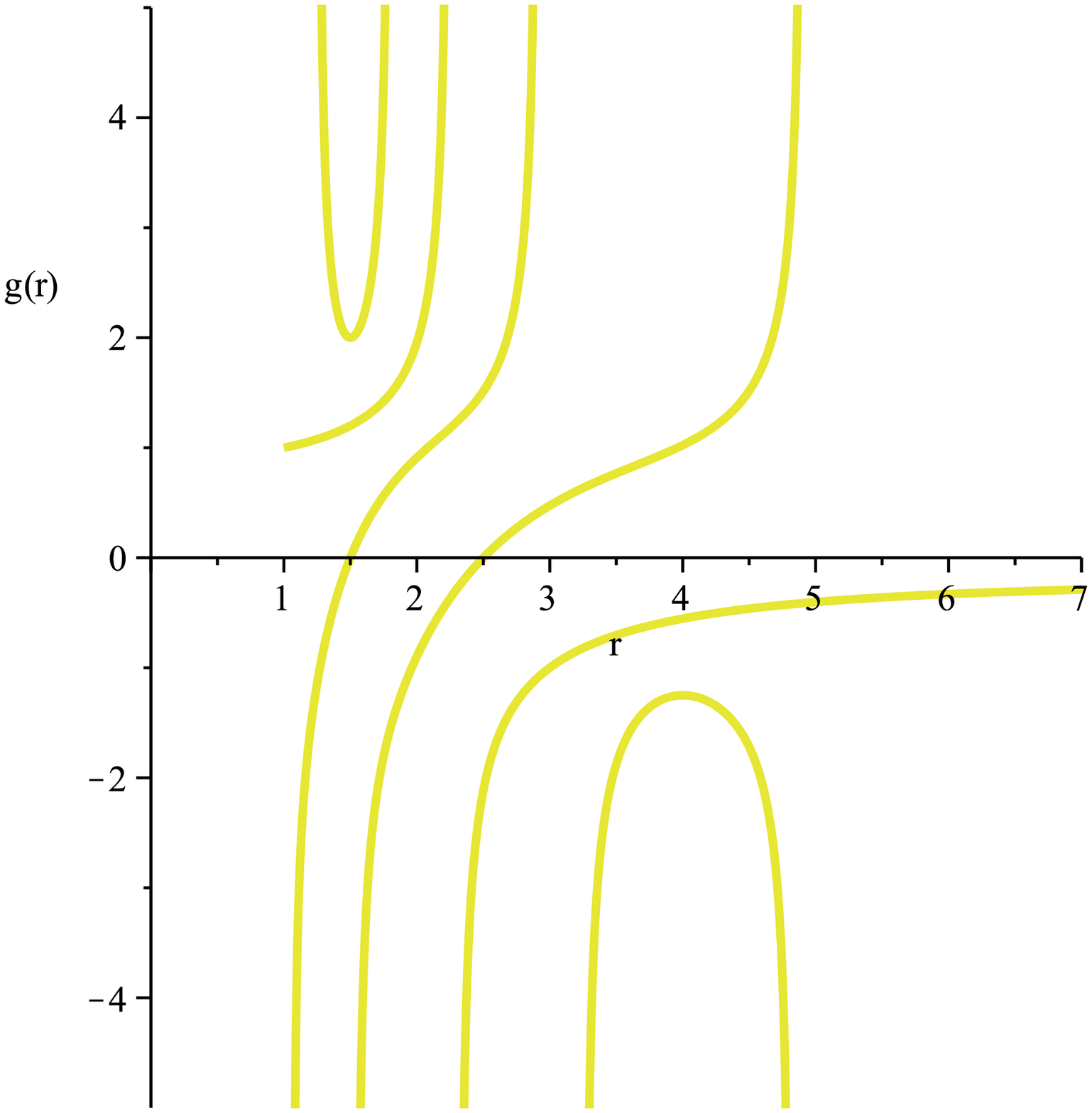}
\caption{Solutions of ODE (\ref{odeeh}) I}
\label{EH1}
\end{minipage}
\hspace{0.5cm}
\begin{minipage}[b]{0.5\linewidth}
\centering
\includegraphics[width=60mm]{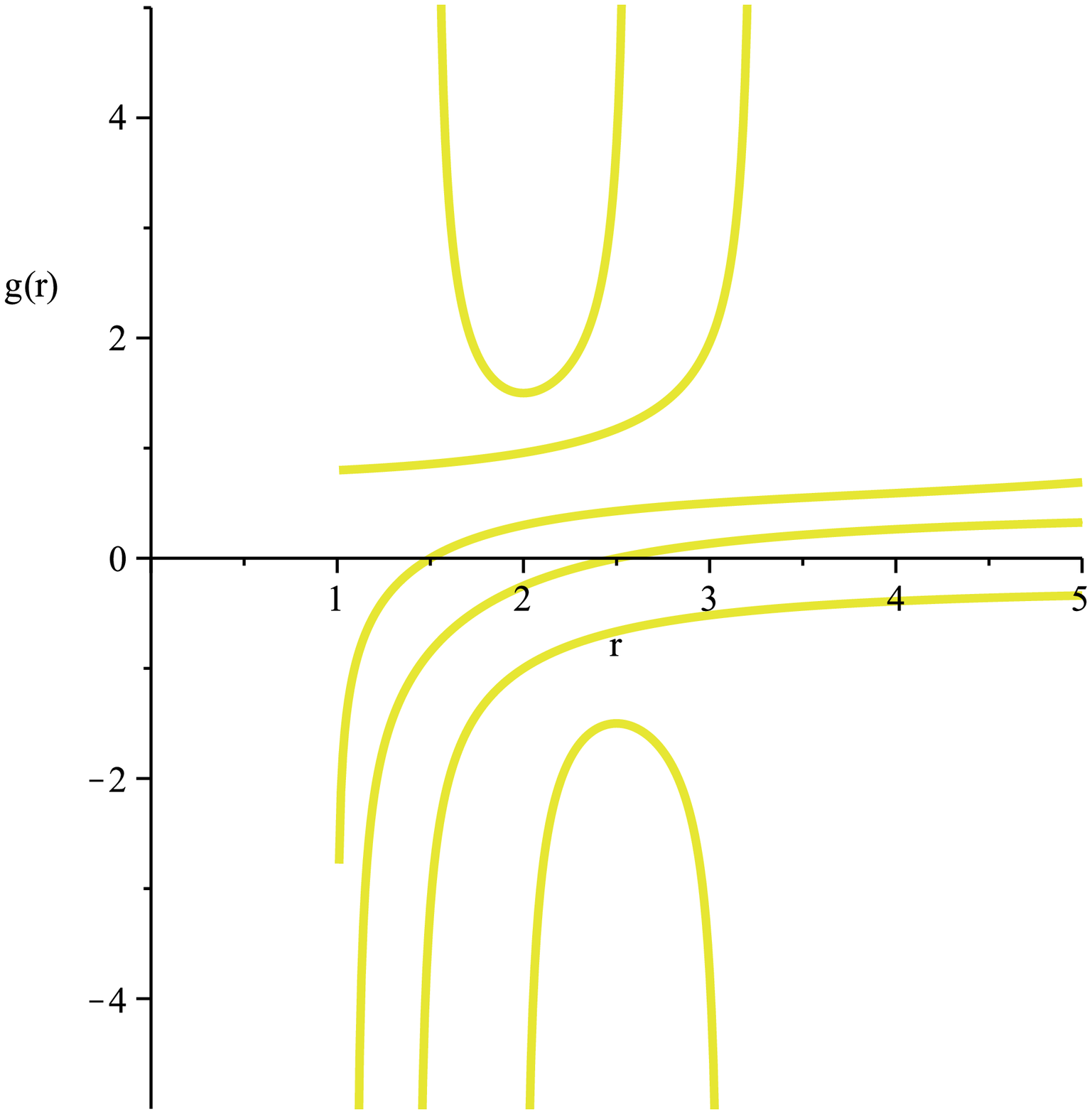}
\caption{Solutions of ODE (\ref{odeeh}) II}
\label{EH2}
\end{minipage}
\end{figure}

Consider $(M_{4}, g_{4})$  to be  the Eguchi-Hanson manifold 
\cite{eguchi-hanson60}, with the metric
\[
g_{4} = \left( 1- \frac{a^4}{r^4} \right)^{-1} dr^2 + \frac{1}{4} r^2 \left(1-\frac{a^4}{r^4} \right) \sigma_3^2 + \frac{1}{4} r^2 ( \sigma_1^2 + \sigma_2^2).
\]
Here $\sigma_i, i=1, 2, 3$ are the left--invariant one--forms on $SU(2)$
\[
 \nonumber
 \sigma_1 + i \sigma_2 = e^{-i\psi} ( d\theta + i \sin \theta d\phi ), \quad \quad \sigma_3 = d\psi + \cos \theta d\phi 
\]
and to obtain the regular metric we take the ranges
\begin{equation}
 \nonumber
 r > a, \quad 0 \leq \theta \leq \pi, \quad 0 \leq \phi \leq 2 \pi, \quad 0 \leq \psi \leq 2 \pi.
\end{equation}
Choose an orthonormal frame
\be
 \label{ehframe}
e^0 = \frac{1}{\sqrt{1-\frac{a^4}{r^4}}} \; dr, \quad  e^1 = \frac{r}{2} \sqrt{1-\frac{a^4}{r^4}} \; \sigma_3,\quad
 e^2 = \frac{r}{2} \; \sigma_2, \quad e^3 = \frac{r}{2} \; \sigma_1.
\ee
Computing the exterior derivatives $d(e^a)$ explicitly we can evaluate 
(\ref{Box12}) and find that it is trivially zero. Furthermore, we know that equations (\ref{PT1}) and (\ref{ASD1}) are equivalent to (\ref{gensol2}). The integrability conditions $d^2H = 0$ imply
\[
 df = 2f \wedge dG, \quad\mbox{where}\quad f=\delta_a e^a
\]
The condition  $dG \neq 0$ implies $\delta_i=0$. Then
\[
 f = \frac{\delta_0 dr}{\sqrt{1-\frac{a^4}{r^4}}},
\]
and $df = 0$. Thus $f \wedge dr = dH \wedge dr = dH \wedge dG = 0$ and consequently $H$ and $G$ depend on $r$ only and satisfy the following relation:
\[
 \frac{dH}{dr} = \frac{\delta_0 e^{2G}}{\sqrt{1-\frac{a^4}{r^4}}}.
\]
Using this in equation (\ref{Box1}) and substituting
$g:= \frac{e^{G}}{\sqrt{\delta_0}}$ yields
\begin{equation}
 \label{odeeh}
 \left( 1- \frac{a^4}{r^4} \right) g'' + \frac{1}{r} \left( 3 + \frac{a^4}{r^4} \right) g' - g^5 = 0.
\end{equation}
The numerical results  (Figures \ref{EH1} and \ref{EH2}, where $a=1$)
indicate that yet again there are no regular functions among the solutions.
Analysing the limits  $r \rightarrow a$ and $r \rightarrow \infty$ we find that
the solution curves either blow up for $r \rightarrow a$ or, if they intersect with the line $r=a$ in the $(r,g)$ plane, 
they will satisfy $g'=(a/4)g^5$.
For the second limit (\ref{odeeh}) tends to $g'' = g^5$ which we 
have investigated in the previous section.
Thus the behaviour for $r \rightarrow \infty$ is determined by Figure \ref{R8}.
In the flat limit $a \rightarrow 0$, in which the Eguchi-Hanson manifold 
becomes $\R^4$, equation (\ref{odeeh})
does not reduce to the one we found for the ansatz over $\R^4$. This is to be expected, since the frame $e^a$ we are working with will not reduce to an integrable coordinate frame even in the flat limit.\\

\subsubsection*{Nonabelian Seiberg--Witten  equations on  
Bianchi II domain wall}
In this Section we shall prove Theorem \ref{theo1}.
Consider the  Gibbons--Hawking \cite{Gibbons:1979zt} class of hyper--K\"ahler metrics characterised by
the existence of a tri--holomorphic isometry. The metric is
given by
\be
\label{GH}
 g_4 = V \left( (dx^1)^2 + (dx^2)^2 + (dx^3)^2 \right) + V^{-1} \left( dx^0 + \alpha \right)^2.
\ee
The function  $V$ and the one--form $\alpha=\alpha_idx^i$ depend on $x^j$
and satisfy
\[
 \ast_3 dV = -d\alpha,
\]
where $\ast_3$ is the Hodge operator on $\R^3$.
Thus the function $V$ is harmonic.
 
Chose the orthonormal frame
\[
 e^0 = \frac{1}{\sqrt{V}} (dx^0 + \alpha), \quad e^i = \sqrt{V} dx^i,
\]
and the dual vector fields $\nabla_0$ and $\nabla_i$. 
In comparison to the Eguchi-Hanson background, for the Gibbons-Hawking case
the equation (\ref{Box12}) is no longer 
trivially satisfied. It only holds if $dG\wedge dV=0$.
Thus, in particular $\nabla_0 G=0$.
The equations (\ref{PT1}) and (\ref{ASD1}) are equivalent to
(\ref{gensol2}).
The integrability conditions force $\delta_0=0$. Setting $w=\delta_i x^i$, 
we can determine $H$ from
the relation
$
dH = \sqrt{V} e^{2G} dw.
$
Thus $H$ and $\sqrt{V} e^{2G}$ are functions of $w$ only. We claim that 
$\sqrt{V} e^{2G} \neq C$ for any constant\footnote{Suppose the opposite. 
Using $V = C^2 e^{-4G}$ in (\ref{Box1}) we find
$
 \p_i \p^i G + \p_i G \p^i G = C^2 \delta_i \delta^i.
$
The Laplace equation on $V$ implies
$
 \p_i \p^i G = 4 \p_i G \p^i G,
$
and
\[
\p_i \p^i G = 4 c^2, \quad \p_i G \;\p^i G = c^2, \quad\mbox{where}\quad c:=\frac{C^2 \delta_i \delta^i}{\sqrt{5}}.
\]
Differentiation of the first relation reveals that all derivatives of $G$ are harmonic. Two partial differentiations of the second relation 
and contracting the indices then yields 
$
| \p_i \p_j G |^2 = 0.
$
This implies $c=0$ and thus $\p_i G = 0$, which rules out this special case.} $C$.
Therefore  $dV \wedge dw = dG \wedge dw = 0$, since $dV \wedge dG = 0$, and we  must have $V := V(w)$, 
$G  := G(w)$. Furthermore $V(w)$ is harmonic, so the potential must be linear in $w$, i.e. without loss of generality 
\[
V = x^3,\quad \alpha=x^2dx^1.
\] 
The resulting metric admits a Bianchi II (also called $Nil$) group of isometries
generated by the vector fields
\[
X_0=\frac{\p}{\p x^0}, \quad X_1=\frac{\p}{\p x^1},\quad
X_2=\frac{\p}{\p x^2}-x^1\frac{\p}{\p x^0}
\]
with the Heisenberg Lie algebra structure
\[
[X_0, X_1]=0,\quad [X_0, X_2]=0, \quad [X_2, X_1]=X_0.
\]
There is also a homothety generated by
\[
D=2x^0\frac{\p}{\p x^0}+x^1\frac{\p}{\p x^1}+x^2\frac{\p}{\p x^2}
+x^3\frac{\p}{\p x^3},
\]
such that 
\[
{\mathcal L}_D g_4=3 g_4.
\]
The conformally rescaled metric $\hat{g}=(x^3)^{-3} g_4$
admits $D$ as as a proper Killing vector.
Thus $\{X_0, X_1, X_2\}$ span the Bianchi II algebra of isometries
of $\hat{g}$ and $\{X_0, X_1, D\}$ span the Bianchi V group of isometries of $\hat{g}$.
Setting $x^3=\exp{(\rho)}$ puts $g_4$ in the form
\[
g_4=e^{3\rho}(d\rho^2+e^{-2\rho}((dx^1)^2+(dx^2)^2)+e^{-4\rho}
(dx^0+x^2dx^1)^2).
\]
This metric is singular at $\rho\rightarrow \pm \infty$ but we claim
that this singularity is only present in an overall conformal factor,
and $g_4$ is a conformal rescaling of a regular homogeneous metric
on a four--dimensional Lie group with the underlying manifold 
${\mathcal H}=Nil\times \R^+$ generated
by the right--invariant vector fields $\{X_0, X_1, X_2, D \}$. To see it, set
\[
\sigma_0=e^{-2\rho} (dx^0+x^2dx^1), \quad
\sigma_1=e^{-\rho} dx^1, \quad \sigma_2=e^{-\rho} dx^2, \quad
\sigma_3=d\rho.
\]
Then
\be
\label{hc}
g_4=e^{3\rho}\hat{g}\quad\mbox{where}\quad \hat{g}=
{\sigma_0}^2+{\sigma_1}^2+{\sigma_2}^2+{\sigma_3}^2,
\ee
and the left--invariant one--forms satisfy
\be
\label{group}
d\sigma_0=2\sigma_0\wedge\sigma_3-\sigma_1\wedge\sigma_2, \quad d\sigma_1=\sigma_1\wedge\sigma_3, 
\quad d\sigma_2=\sigma_2\wedge\sigma_3,\quad
d\sigma_3=0.
\ee
Thus the metric $\hat{g}$ is regular.

In \cite{Gibbons:1998ie} the singularity of $g_4$ at $\rho=-\infty$
has been interpreted as a single side domain wall in
the space--time
\[
M_4\times \R^{p-3,1}
\]
with its product metric. This domain wall is a $p$--brane: either
a nine--brane of 11D super gravity if $p=6$ or a three--brane
of the $4+1$ dimensional space--time $g_4-dt^2$. In all cases
the direction $\rho$ is transverse to the wall. In the approach
of \cite{Gibbons:1998ie} the regions $x^3>0$ and $x^3<0$ are identified.
In this reference it is argued that $(M_4, g_4)$ with such identification
is the approximate form of a regular metric constructed in
\cite{Kobayashi90} on a complement of a smooth cubic curve in $\CP^2$.
\vskip5pt

\begin{figure}[ht]
\begin{minipage}[b]{0.5\linewidth}
\centering
\includegraphics[width=60mm]{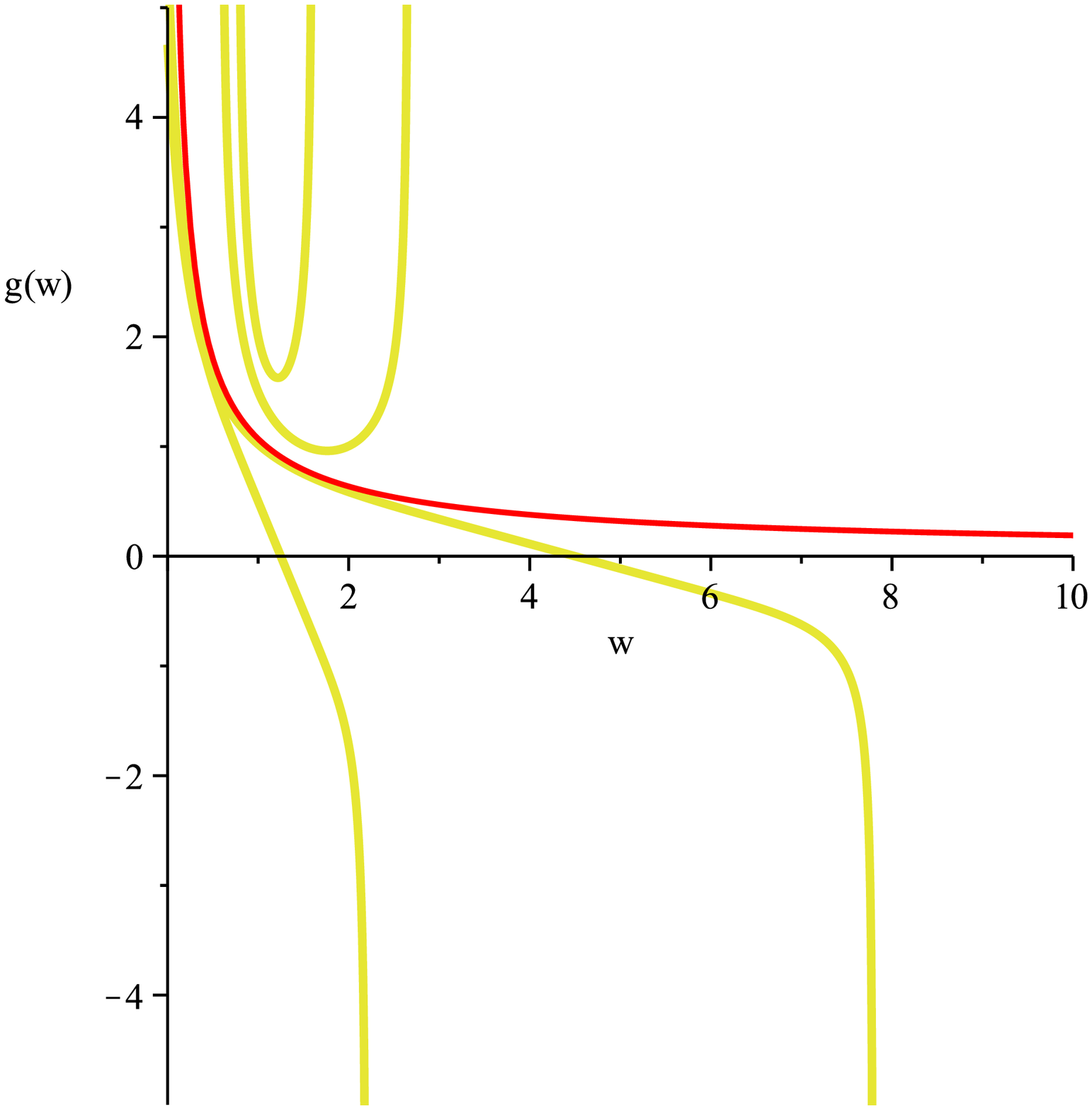}
\caption{Solutions of ODE (\ref{odegh}) I}
\label{GH1}
\end{minipage}
\hspace{0.5cm}
\begin{minipage}[b]{0.5\linewidth}
\centering
\includegraphics[width=60mm]{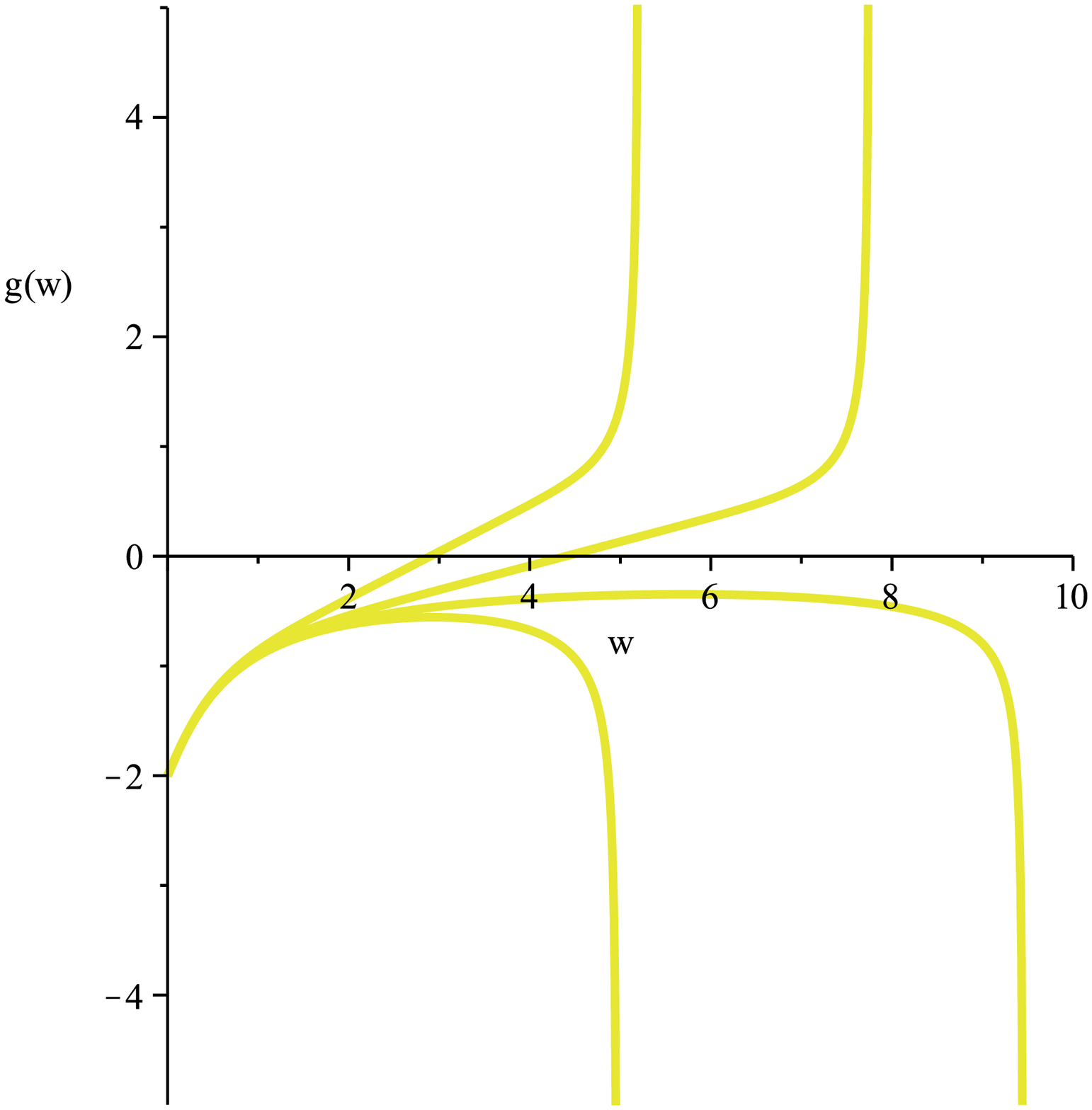}
\caption{Solutions of ODE (\ref{odegh}) II}
\label{GH2}
\end{minipage}
\end{figure}

 Using this linear potential $V=w=x^3$ in (\ref{Box1}) and setting $g(w):= e^{G(w)}$
yields
\[
 \label{odegh}
g'' - wg^5 =0.
\]
This equation changes its character as $w$ changes from positive to negative sign, we find infinitely many singularities for $G(w)$ for $w<0$. We thus focus on the region $w>0$, which is in agreement with the identification of these two regions proposed by \cite{Gibbons:1998ie}. Numerical plots for solutions of this equation are given in Figures \ref{GH1} and \ref{GH2}. One explicit solution is given by
\be
\label{sol11}
 g(w) = \pm \frac{1}{2}  \sqrt[4]{21} w^{-\frac{3}{4}}.
\ee
If we choose $w= x^3$, the curvature for this solution blows up like $(x^3)^{-3}$. This is singular
only on the domain wall.
Explicitly, the solution (\ref{sol11}) gives
\[
G=-\frac{3}{4}\rho+\frac{1}{4}\ln{21}-\ln{2}, \quad
H=-\frac{\sqrt{21}}{3} G.
\]
and
\begin{eqnarray}
\label{gauge_f}
A&=&\frac{3}{4}(\sigma_2\otimes T_1-\sigma_1\otimes T_2+\sigma_0\otimes T_3),\quad \Phi=-\frac{\sqrt{21}}{3} A,\\
F&=&\Big(\frac{9}{16}\sigma_0\wedge\sigma_1
+\frac{3}{4}\sigma_2\wedge\sigma_3\Big)
\otimes T_1+\Big(\frac{9}{16}\sigma_0\wedge\sigma_2-\frac{3}{4}
\sigma_1\wedge \sigma_3\Big)\otimes T_2\nonumber\\
&&+\Big(\frac{3}{2}\sigma_0\wedge\sigma_3
-\frac{3}{16}\sigma_1\wedge\sigma_2\Big)\otimes T_3.\nonumber
\end{eqnarray}
 We claim that $(A, \Phi)$ is a regular solution to the non--abelian Seiberg--Witten equations
on the Lie group corresponding to the Lie algebra (\ref{group})
with its left--invariant metric $\hat{g}$ given by (\ref{hc}).
To justify this claim, we need to consider the invariance
of the non--abelian Seiberg--Witten equations under the conformal
rescalings of the underlying metric.  The first two equations
(\ref{FPhi}) and (\ref{DPhi})
are clearly invariant, which follows from the conformal invariance of the Hodge operator acting on two--forms
in four dimensions. The third equation (\ref{DivPhi}) is not invariant 
in general, but it still holds in our case with $g_4$ replaced
by $\hat{g}_4$, as the conformal factor depends only on $\rho$ and
$d\rho\wedge *_4\Phi=0$ for the Higgs fields (\ref{gauge_f}).
We should stress that this solution does not lift to a solution
of Yang--Mills equations in eight dimensions, as the product
metric $\hat{g}_4+\tilde{g}_4$ on ${\mathcal H}\times \widetilde{R}^4$
is not $Spin(7)$.

\section{Conclusions and Outlook}
In this paper
we have used the identification of $\R^8$ with $\R^4\times \R^4$, or
the curved analogue when one of the $\R^4$ factors is replaced
by a hyper--K\"ahler four manifold $(M_4, g_4)$ to 
construct explicit solutions
of the `self--duality' equations in eight dimensions with a gauge group $SU(2)$. The solutions all admit four dimensional symmetry group
along the $\R^4$ factor, and thus they give rise
to solutions of the non--abelian Seiberg--Witten equations on $M_4$.

 We have analysed three cases, where $M_4$ is $\R^4$ 
with the flat metric, the Eguchi--Hanson gravitational instanton, and
finally the cohomogeneity one hyper--K\"ahler metric with Bianchi
II group acting isometrically with three--dimensional orbits.
In this last case the singularity of the gauge field
is regular on a conformally rescaled four--manifold. Alternatively, the singularity is present only on a domain wall in the space--time with the metric
$g_4-dt^2$.
\vskip5pt
The symmetry reduction to four dimensions was based on the holonomy reduction $SU(2)\times SU(2)\subset Spin(7)$. An analogous reduction 
from $\R^{8}$ with split signature metrics may provide a source of Lorentz  invariant gauged solitons in $3+1$ dimensions. 
Moreover, there are other special realisations
of $Spin(7)$ in terms of Lie groups $G_2$, $SU(3)$ and $SU(4)$. 
Each realisation leads to some symmetry reduction
\cite{harland,popov2},
and picks a preferred
gauge group, where the ansatz analogous to (\ref{ansatz}) can be made.
\vskip5pt
Witten \cite{Witten} considered a complex--valued connection 
${\mathcal A}=A+i\Phi$ on bundles over four--manifolds of the form
$M_4=\R\times M_3$ with the product metric $g_4=dw^2+g_3$, where
$(M_3, g_3)$ is a three--dimensional Riemannian manifold. He showed that
the gradient flow equation
\[
\frac{d\mathcal A}{dw}=-*_3\frac{\delta {\mathcal I}}{\delta \bar{\mathcal{A}}}
\]
for the holomorphic Chern--Simons functional $\mathcal{I}$
yields equations 
(\ref{FPhi}) and (\ref{DPhi}).
 In this setup neither $A$ nor $\Phi$ have a 
$dw$ component. 

 The example (\ref{static_sol}) fits into this framework:
$g_3$ is the flat metric on $\R^3$, and the corresponding ODE
is the reduction of the gradient flow equations.
In all other examples in our paper the underlying four manifold
is also of the form $M_4=\R\times M_3$, where $M_3$ is a three dimensional 
Lie group with left--invariant one--forms $\sigma_i$. Moreover
in all cases there exists a gauge such that neither $A$ nor $\Phi$ have
components in the $\R$--direction orthogonal to the group orbits.
However the Riemannian metric $g_4=dw^2+h_{ij}(w)\sigma_i\sigma_j$ on $M_4$
is not a product metric unless $h_{ij}$ does not depend on $w$.
It remains to be seen whether the gradient flow formulation of the non--abelian
Seiberg--Witten equations can be achieved in this more general setup.          
\section*{Appendix}
\setcounter{equation}{0}
\appendix
\def\theequation{\thesection{A}\arabic{equation}}
{\bf Proof of Proposition \ref{prop1}.}
Rewrite equations (\ref{F+=0}) using the two-forms $\sigma$ and $\tilde{\sigma}$:
\begin{eqnarray}
 \label{proofI} \ast [\sigma \wedge (F - \frac{1}{2} [\Phi, \Phi] )] = \sigma^{ab} \left( F_{ab} - \Phi_a \wedge \Phi_b \right) &=&0, \\
 \label{proofII} \ast (\tilde{\sigma} \wedge [D\Phi]) = - \tilde{\sigma}^{ab} D_a \Phi_b &=&0, \\
 \label{proofIII} D^a \Phi_a &=&0.
\end{eqnarray}
Now, substituting  (\ref{ansatz}) and using (\ref{sigmaId}) 
in equation (\ref{proofI}) yields
\begin{eqnarray*}
0&=&\frac{1}{2} \sigma^{ab} \left( F_{ab} - \frac{1}{2} \left[ \Phi_a, \Phi_b \right] \right) =\\
&=& \frac{3}{4} \nabla_a \nabla^a G + \sigma_{ac} \nabla^a \nabla^c G + \sigma_{cd} \nabla^d G \sigma^{ab} d(e^c)_{ab} + \frac{3}{4}| \nabla G |^2 - \frac{3}{4} | \nabla H |^2.
\end{eqnarray*}
The term $\sigma_{cd} \nabla^d G \sigma^{ab} d(e^c)_{ab}$ decomposes as
\[
 \sigma_{cd} \nabla^d G \sigma^{ab} d(e^c)_{ab} = \frac{1}{4} \left[ {C^a}_{da} + {\epsilon_{da}}^{bc} {C^a}_{bc} \right] \nabla^d G \; \ID + {\epsilon_{ea}}^{bc} {C^a}_{bc} \nabla^d G {\sigma^e}_d.
\]
The closure condition  $d\sigma = 0$ yields
$
 \sigma_{a \left[ b \right.} {C^a}_{\left. cd \right] } = 0,
$
which is a system of 12 linear equations. 
These equations imply the four relations
$
 \label{closedID}
 {\epsilon_{da}}^{bc} {C^a}_{bc} = 2 {C^a}_{da}.
$
Then the identity-valued part of (\ref{proofI}) becomes
\[
 \frac{3}{4} \nabla_a \nabla^a G + \frac{3}{4} {C^a}_{ba} \nabla^b G + \frac{3}{4} | \nabla G |^2 - \frac{3}{4} | \nabla H |^2 = 0
\]
The first two terms of these combine to give 
$\Box G$, as can be seen by computing
\begin{eqnarray*}
 \Box G &=& \ast d \ast d G = \ast d ( \frac{1}{3!} \epsilon_{abcd} \nabla_a G e^b \wedge e^c \wedge e^d ) \\
        &=& \ast ( \nabla_a \nabla^a G + {C^b}_{ab} \nabla^a G ) = ( \nabla_a \nabla^a + {C^b}_{ab} \nabla^a ) G.
\end{eqnarray*}
The other components of (\ref{proofI}) are given 
by\footnote{Using the spinor decomposition \cite{Dbook} \[
{C^a}_{bc}={\varepsilon^{A'}}_{B'}{\Gamma^A}_{BCC'}+
{\varepsilon^{A}}_{B}{\Gamma^{A'}}_{B'CC'}
\]
with the anti--self--duality conditions $d\sigma=0$ equivalent to
${\Gamma^{A'}}_{B'CC'}=0$ gives
\[
{\Gamma^{AB}}_{AC'}\sigma^{C'B'}\nabla_{BB'} G=0,
\]
where $\sigma^{A'B'}=\sigma^{(A'B')}$ and $\sigma^{ab}=\sigma^{A'B'}\varepsilon^{AB}$.
Thus the three--dimensional distribution  ${\Gamma^{AB}}_{A(C'}\nabla_{B')B}$ is integrable and $G$ is in its
kernel.}
\[
 \left( {\epsilon_{ea}}^{bc} {C^a}_{bc} \sigma^{ed} - \sigma^{ab} {C^d}_{ab} \right) \nabla_d G = 0.
\]
We now move to equation (\ref{proofII}),
\begin{eqnarray*}
 \tilde{\sigma}_{ab} \left( D^a \Phi^b \right)
    &=& \tilde{\sigma}_{ab} \left( \nabla^a \Phi^b + A^a \Phi^b - \Phi^b A^a \right)\\
    &=& \tilde{\sigma}_{ab} \sigma^{bc} \nabla^a \nabla_c H + 2 \tilde{\sigma}_{ab} \sigma^{ad} \sigma^{bc}  \nabla_{\left( c \right.} G \nabla_{\left. d \right)} H \\
    &=& \tilde{\sigma}_{ab} {\sigma^b}_c \left(  \nabla^a \nabla^c H - 2 \nabla^a H \nabla^c G \right). 
\end{eqnarray*}
Here we had to explicitly evaluate and symmetrise a product of three $\sigma$-matrices to obtain the last line. And finally, for equation (\ref{proofIII}) we obtain
\begin{eqnarray*}
 D_a \Phi^a &=& \left( \nabla_a \Phi^a + \left[ A_a, \Phi^a \right] \right) =\\
           &=& \nabla_a \left( \sigma^{ab} \nabla_b H \right) + \sigma_{ab} {\sigma^a}_c \nabla^b G \nabla^c H - \sigma_{ac} {\sigma^a}_b \nabla^b G \nabla^c H \\
          &=& \sigma_{ab} \left( \nabla^a \nabla^b H - 2 \nabla^a G \nabla^b H \right) = 0.
\end{eqnarray*}
\koniec

\end{document}